\documentclass[aps,prd,superscriptaddress,showpacs,preprintnumbers]{revtex4}

\usepackage{color}
\usepackage{epsfig}
\usepackage{graphicx,graphics}
\usepackage{float}
\usepackage{amsmath}
\usepackage{amsfonts}
\usepackage{amssymb}
\usepackage{multirow}

\newcommand{\bi}{\begin{itemize}}
\newcommand{\ei}{\end{itemize}}

\usepackage{bm}

\newcommand{\nl}{\newline}
\newcommand{\beq}{\begin{equation}}
 \newcommand{\eeq}{\end{equation}}

 \newcommand{\be}{\begin{eqnarray}}
 \newcommand{\ee}{\end{eqnarray}}
 \newcommand{\nee}{\nonumber\end{eqnarray}}

  \newcommand{\bc}{\begin{center}}
 \newcommand{\ec}{\end{center}}

\def\kt{k_\perp}

\def\pp{p_\perp}

\def\avk{\langle k_\perp ^2\rangle}
\def\avp{\langle p_\perp ^2\rangle}
\def\avPT{\langle P_T^2\rangle}

\def\S{_{_S}}

\def\C{_{_C}}

\def\BM{_{_{B\!M}}}

\def\xb{x_{_{\!B}}}
\def\1{_{_{C\!1}}}
\def\2{_{_{S1}}}

  \def\a               {\alpha}
\def\b               {\beta}

\def\g              {\gamma}

\begin{document}

\author{E. Christova}
\email {echristo@inrne.bas.bg} \affiliation{Institute for Nuclear
Research and Nuclear Energy, Bulgarian Academy of Sciences,
Tzarigradsko chauss\'{e}e 72, 1784 Sofia, Bulgaria}
\author{D. Kotlorz}
\email {dorota@theor.jinr.ru}
\affiliation{Department of Physics, Opole University of Technology,
45-758 Opole, Proszkowska 76, Poland}
\affiliation{Bogoliubov Laboratory of Theoretical Physics, JINR,
141980 Dubna, Russia}\textcolor[rgb]{0.00,0.00,1.00}{}
\author{E. Leader}
\email {e.leader@imperial.ac.uk} \affiliation{ Imperial College
London, London SW7 2AZ, United Kingdom}
\title{Towards a model independent extraction  of the Boer-Mulders  function}

\begin{abstract}
At present, the Boer-Mulders  function for a given quark flavour
has been  extracted from data on semi-inclusive deep inelastic scattering
 using  the simplifying, but theoretically inconsistent,  assumption that it is proportional to
the Sivers function for each quark flavour.
In this paper, using the latest semi-inclusive deep inelastic  COMPASS deuteron data on
the $\langle\cos\phi_h\rangle$ and   $\langle\cos 2 \phi_h\rangle$ asymmetries we extract the
collinear $\xb$-dependence of the Boer-Mulders function
for the sum of the valence quarks $Q_V=u_V + d_V$ in an essentially  model independent way,
and find a significant disagreement with the published results.
Our analysis also yields interesting information on the transverse momentum dependence
of the unpolarized quark distribution and fragmentation functions.
\end{abstract}

\pacs{...}

\date{\today}

\maketitle

The Boer-Mulders (BM) function is an essential element in describing the internal structure of
the nucleon.  In a nucleon of momentum $\bf P$, and for
a quark with transverse momentum ${\bf k}_\perp$, the BM function
measures the difference between the number density of quarks polarized
parallel and anti-parallel to $ {\bf ( P \times k_\perp)}$. Past attempts to extract it from experiment
were hindered by the scarcity of data and  made the theoretically inconsistent simplifying assumption
\cite{Barone:2009hw, Barone:2008tn}
that for each quark flavour, it  is proportional to the better known Sivers function.

In this paper, we show that the new COMPASS data on the unpolarized
$\langle\cos\phi_h\rangle$ and $\langle\cos 2\phi_h\rangle$ asymmetries in
semi-inclusive deep inelastic scattering (SIDIS) reactions for producing a hadron $h$ and and its antiparticle $\bar h$
at azimuthal angle $\phi_h$, allows an essentially model independent extraction of the BM function.

As explained in \cite{Christova:2000nz} and \cite{Christova:2015jsa}  there is a great advantage in studying
difference asymmetries $A^{h-\bar h}$, effectively $ A^h -A^{\bar h}$, since both for the collinear and transverse momentum dependent
(TMD) functions, only the flavour non-singlet  valence quark parton densities (PDFs) and fragmentation functions
(FFs) play a role and the gluon does not contribute.  On a deuteron target an additional simplification occurs that
independently of the final hadron,  only the sum of the valence-quark TMD functions  $Q_V=u_V+d_V$ enters.
In this paper we use SIDIS COMPASS data on a deuteron target, \cite{Adolph:2014pwc},
and determine the BM TMD function only for $Q_V$, but with essentially no model assumptions.

The  unpolarized TMD functions for $Q_V$ are parametrized
in the standard way~\cite{Anselmino:2011ch, Christova:2014gva}:
\be
f_{Q_V/p}(\xb,k_\perp^2 ,
Q^2)&=&Q_V(\xb,Q^2)\,\frac{e^{-k_\perp^2/\avk}}{\pi\avk}\label{fq}
\ee
 and
 \be
D_{h/q_V}(z_h,p_\perp^2 ,Q^2) =D_{q_V}^h(z_h,Q^2)
\,\frac{e^{-p_\perp^2/\avp }}{\pi\avp },\label{Dq}
\ee
where $Q_V(\xb,Q^2)$ is the sum of the collinear valence-quark  PDFs:
 \be
 Q_V(\xb,Q^2)= u_V(\xb,Q^2) +d_V(\xb,Q^2)
  \ee
  and $D_{q_V}^h(z_h,Q^2)$ are the valence-quark collinear  FFs:
   \be
  D_{q_V}^h(z_h,Q^2)=D_{q}^h(z_h,Q^2)-D_{\bar q}^h(z_h,Q^2),
  \ee
   $\avk $ and $\avp$ are
parameters extracted from a study of the multiplicities in unpolarized SIDIS.
There is some controversy in the literature about their values. This study will suggest a resolution
of the problem.

The BM function is parametrized in a similar way:
\beq
\hspace*{-.5cm}  \Delta  f^{Q_V}\BM (\xb,\kt ,Q^2) \!=\! \Delta
f^{Q_V}\BM (\xb,Q^2)\; \sqrt{2e}\,\frac{\kt}{M\BM } \;
\frac{e^{-\kt^2/\avk\BM  }}{\pi\avk },\label{BM-Siv_dist1}
\eeq
with
\beq
\Delta
f^{Q_V}\BM (\xb,Q^2)\!=\! 2\,{\cal N}\BM ^{Q_V}(\xb)\,Q_V(\xb,Q^2).
\label{BM-Siv_dist2}
 \eeq
Here the ${\cal N}^{Q_V}\BM (\xb)$ is an unknown function and $M\BM $,  or equivalently $\avk \BM$:
\be
 \avk \BM = \frac{\avk  \, M^2\BM }{\avk  + M^2 \BM },
 \ee
is an unknown  parameter.

Since the asymmetries under study involve a product of the BM parton density
and the Collins FF, one requires also the transverse momentum dependent Collins function:
\beq
 \Delta^N  D_{h/u_V\uparrow}(z_h,\pp ,Q^2) \!=\!\Delta^N  D_{h/u_V\uparrow}(z_h,Q^2)\,
\sqrt{2e}\,\frac{\pp}{M\C} \; \frac{e^{-\pp^2/\avp\C }}{\pi\avp}\,,
\eeq
where
\beq
\Delta^N D_{h/u_V\uparrow}(z_h,Q^2)\!=\!2\,{\cal N}^{h/\!u_V}\C
(z_h)\,D_{u_V}^h(z_h,Q^2) .
\eeq
The  quantities  ${\cal N}^{h/\!u_V}\C (z_h)$ and $M\C$,
or equivalently $ \avp\C$:
\be
 \avp\C  =\frac{\avp  \, M\C ^2}{\avp  +M\C ^2}\,,
\label{Coll-frag2}
 \ee
are known from studies of the azimuthal correlations of pion-pion,
pion-kaon and kaon-kaon pairs produced in $e^+e^-$ annihilation: $e^+e^-\to h_1h_2+X$ and the $\sin (\phi_h +\phi_S)$
asymmetry in polarized SIDIS \cite{Anselmino:2008jk, Anselmino:2015sxa, Anselmino:2015fty}.

Besides the BM-Collins contributions to the $\langle\cos\phi_h\rangle$
and $\langle\cos 2\phi_h\rangle$ unpolarized asymmetries,
there exists also a contribution known as the Cahn effect, which involves only the collinear
unpolarized PDFs and FFs.

In \cite{Christova:2017zxa} we showed that for the range of the COMPASS data,
evolution effects can be safely neglected, leading to simplified expressions for the
$\langle\cos\phi_h\rangle$ and $\langle\cos 2\phi_h\rangle$ asymmetries.

In the following the measured asymmetries, denoted $A_{UU}^{\cos  \phi_h}$ and $A_{UU}^{\cos  2\phi_h}$ correspond to the
definitions used in the COMPASS paper~\cite{Adolph:2014pwc}.
$[\,$Note that several different definitions \cite{DAlesio:2007bjf} of these asymmetries exist in the literature,
some of them even in COMPASS publications \cite{Bradamante:2007ex}$\,]$.
They are related to the theoretical functions via:
    \be
  A_{UU}^{\cos  \phi_h, h-\bar h}&=&\sqrt \frac{\avk}{\langle Q^2\rangle(x_B)}\,
  \left\{{\cal N}\BM^{Q_V}(\xb)\,{\cal C}\BM^h +{\cal C}_{Cahn}^h\right\},   \label{A1}\\
   A_{UU}^{\cos  2\phi_h, h-\bar h}&=&\left\{{\cal N}\BM^{Q_V}(\xb)\,
   \hat {\cal C}\BM^h + \frac{\avk}{\langle Q^2\rangle(x_B)}\,\hat {\cal C}^h_{Cahn}\right\},\label{A2}
    \ee
where $\langle Q^2\rangle (\xb )$ is some mean value of $Q^2$ for each $\xb$-bin and
the coefficients ${\cal C}\BM$, ${\cal C}_{Cahn}$, $\hat {\cal C}\BM$ and $\hat {\cal C}_{Cahn}$
are dimensionless constants given by integrals over
various products of  the unpolarized  or Collins  FFs and, crucially,
whose values depend on  the parameters $\avk$, $\avp$, $M\BM$ and $M\C$.
For a finite range of integration over $P_T^2$, corresponding to the experimental
kinematics, $a\leq P_T^2\leq b$, they are given by the expressions:
\be
{\cal C}_{Cahn}^h &=&
-2\,\frac{\int dz_h\,z_h\,[D_{q_V}^{h}(z_h)]S_1(a,b; \avPT)/(\eta + z_h^2)^{1/2}}{\int dz_h\, [D_{q_V}^{h}(z_h)]\,S_0(a,b; \avPT)}
\label{Cahn} \\
{\cal C}\BM^h &=& 4e\,\frac{\lambda\BM^2\lambda\C^2}{M\BM M\C}\,\avp \nonumber \\
&&\times\frac{\int dz_h\,[\Delta^N D_{{q_V}\!\uparrow}^{h}(z_h)]\,
[\,z_h^2\lambda\BM\,S_3(a,b,\avPT\BM)+\,(\eta\lambda\C-z_h^2\lambda\BM)\,S_1(a,b;\avPT\BM)]/(z_h^2\lambda\BM+\eta\lambda\C)^{3/2}}
{\int dz_h\, [D_{q_V}^{h}(z_h)]\,S_0(a,b; \avPT)}\label{BM}
\\
\hat {\cal C}^h_{Cahn} &=&
\frac{2\int dz_h \left(z_h^2/[\eta +z_h^2]\right)\,[D_{q_V}^{h}(z_h)]\,S_2(a,b;\avPT\BM )}{\int dz_h\,
[D_{q_V}^{h}(z_h)]\,S_0(a,b; \avPT)}\label{hatC}\\
\hat {\cal C}\BM^h &=&
-2e\,\frac{\lambda\BM^2\lambda\C^2}{M\BM M\C}\,\avp\;
\frac{\int dz_h\,[z_h\,\Delta^N
D_{{q_V}\!\uparrow}^{h}(z_h)]/(z_h^2\lambda\BM+\eta\lambda\C)\,S_2(a,b; \avPT)}
{\int dz_h\, [D_{q_V}^{h}(z_h)]\,S_0(a,b; \avPT)}\label{hatBM}\ee
where, with $\tau = \textrm{either}\,\, \avPT\,\, \textrm{or} \,\,\avPT\BM $,
\be
 S_n(a,b; \tau)&=&\int_a^b\,
dP_T^2\,P_T^n e^{-P_T^2/\tau}/\tau^{1+n/2}\, .
 \ee
Here $[D_{q_V}^h]$ and $ [\Delta^N D_{{q_V}\!\uparrow}^{h}(z_h)]$ are combinations of the collinear and Collins FFs:
\be
[D_{q_V}^{h}(z_h,Q^2)] &=& e_u^2\,D_{u_V}^h+e_d^2\, D_{d_V}^h,
\ee
\be
[\Delta^N D_{{q_V}\!\uparrow}^h (z_h,Q^2)] &=&
e_u^2\,\Delta^N D_{{u_V}\!\uparrow}^h +e_d^2\,\Delta^N D_{{d_V}\!\uparrow}^h \label{D2}
\ee
 and
   \be
    \eta =
\frac{\avp}{\avk},\quad \lambda\C  &=&\frac{ M\C ^2}{\avp +M\C
^2},\quad
 \lambda\BM  =\frac{ M\BM ^2}{\avk  +M\BM ^2}\cdot \label{CBM}
 \ee
\\

As mentioned, there is some controversy as to the values of these parameters, with a wide range of values given in literature.
The coefficients ${\cal C}_{Cahn}$, ${\cal C}\BM$, $\hat {\cal C}_{Cahn}$,
$\hat {\cal C}\BM$ are given in Table~1, grouped together in Sets corresponding to the values of these parameters,  with $\rho=-{\cal C}\BM/\hat {\cal C}\BM$.\\
\begin{table}[h]
\begin{tabular}{|c|c|c|c|c|c|c|c|c|c|}
\hline SET & $~\avk~$ & $~\avp~ $ &$~M\BM^2~$  &$~M\C^2~$
 & ${\cal C}_{Cahn}$ & $~{\cal C}\BM~$ & $\hat {\cal C}_{Cahn}$   & $~\hat {\cal C}\BM~$ & $~~\rho~~$ \\ \hline
I    & 0.18 & 0.20 & 0.34 & 0.91 & -0.68 & 2.1  & 0.31 & -0.47  & 4.4 \\
II   & 0.18 & 0.20 & 0.19 & 0.91 & -0.68 & 1.8 & 0.31 & -0.40  & 4.4 \\
III  & 0.25 & 0.20 & 0.34 & 0.91 & -0.77 & 1.9 & 0.38 & -0.49  & 3.8 \\
IV   & 0.25 & 0.20 & 0.19 & 0.91 & -0.77 & 1.4 & 0.38 & -0.39  & 3.7 \\
V    & 0.57 & 0.12 & 0.80 & 0.28 & -1.2  & 0.89 & 0.84 & -0.50  & 1.8 \\ \hline
\end{tabular}
\caption {${\cal C}_{Cahn}$, ${\cal C}\BM$, $\hat {\cal C}_{Cahn}$,
$\hat {\cal C}\BM$ and $\rho$ calculated for different sets of
$\avk$, $\avp$, $M\BM^2$  ($M\BM^2=M\S^2$ assumed) and $M\C^2$.
The parametrizations for the collinear FFs are from AKK'2008 \cite{Albino:2008fy}, and for Collins functions
-- for sets I -- IV  -- from \cite{Anselmino:2008jk} and \cite{Anselmino:2015fty}, and for set V --
from \cite{Anselmino:2015sxa}  and \cite{Anselmino:2015fty}.
The integrations  are according to COMPASS kinematics:
 $0.01\leq P_T^2\leq 1 \,GeV^2$ and $0.2\leq z_h\leq 0.85$ \cite{Adolph:2014pwc}.}
\end{table}
\begin{figure}[H]
\begin{center}
\includegraphics[scale=0.3]{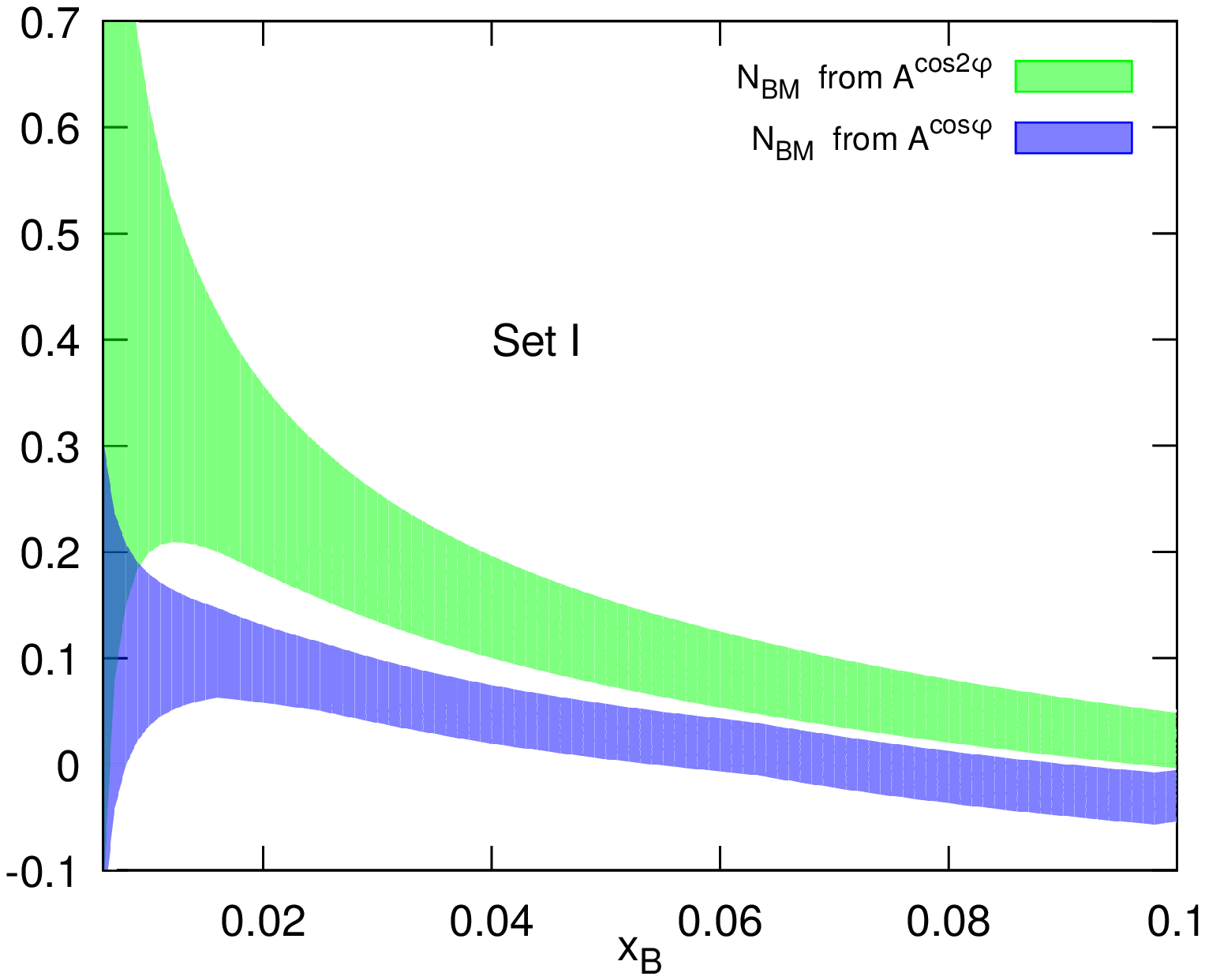}~~~\includegraphics[scale=0.3]{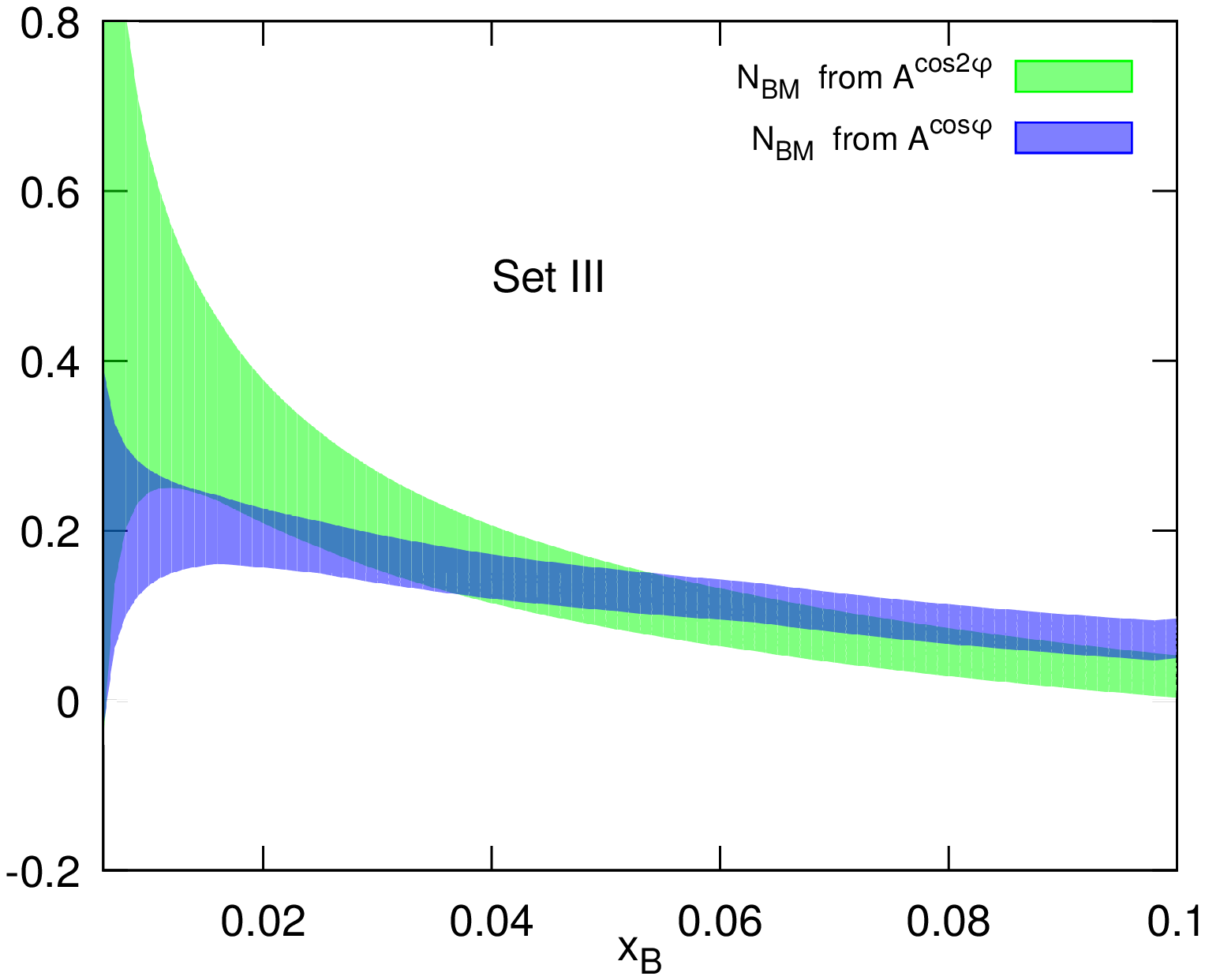}~~~\includegraphics[scale=0.3]{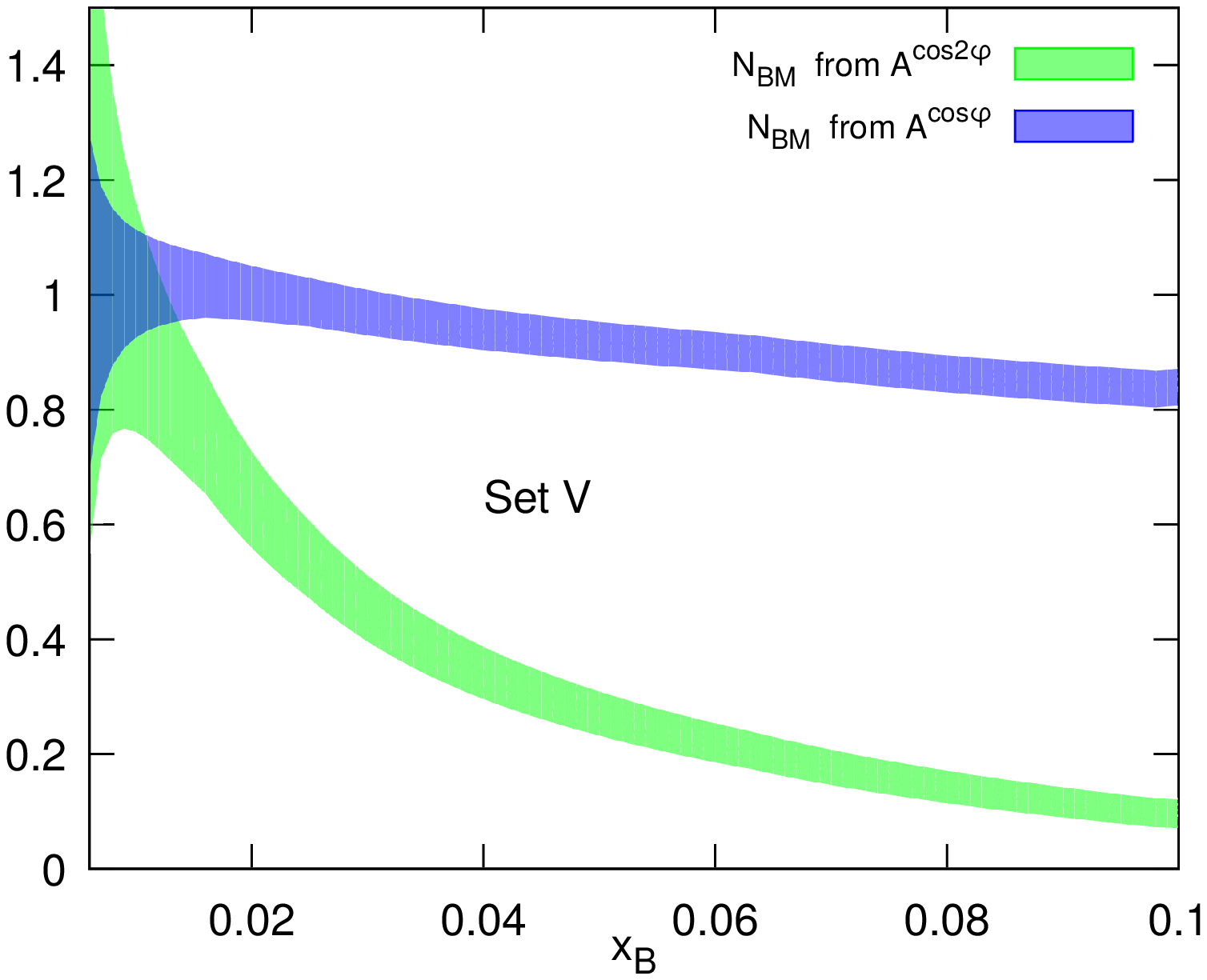}
\end{center}
\caption
{${\cal N}\BM^{Q_V}(\xb)$ extracted from the difference asymmetries, Eqs.~(\ref{A1}) and (\ref{A2}),
using different sets of parameters of Table~I. Plots for Sets. II and IV overlap with those for Sets. I and III, respectively.}
\label{fig:1}
\end{figure}
We form the type of difference  asymmetries
$A^{h^+-h^-}_J$ advocated in \cite{Christova:2017zxa}
from the corresponding usual asymmetries $A_j^{h^+}$ and $A_j^{h^-}$ for
positive  and negative charged hadron production measured in COMPASS \cite{Adolph:2014pwc} via the relation \cite{Alekseev:2007vi}:
 \be
    A_J^{h^+-h^-} = \frac{1}{1-r} \left( A_J^{h^+} - r A_J^{h^-}\right),\qquad J=\langle\cos\phi_h\rangle, \langle\cos 2\phi_h\rangle .
    \label{eq.Diff}
\ee
Here $r$ is the ratio of the unpolarized $\xb$-dependent SIDIS
cross sections for production of  negative and positive hadrons
$ r = \sigma^{h^-}(\xb ) / \sigma^{h^+}(\xb )$ measured in the same kinematics \cite{Alekseev:2007vi}.
In practice we construct the difference asymmetries using smooth fits  to the
data on the usual asymmetries and to the ratio $r$. For $ \langle Q^2\rangle(\xb)$
we perform a linear interpolation of the  COMPASS data points.

\begin{figure}[H]
\begin{center}
\includegraphics[scale=0.3]{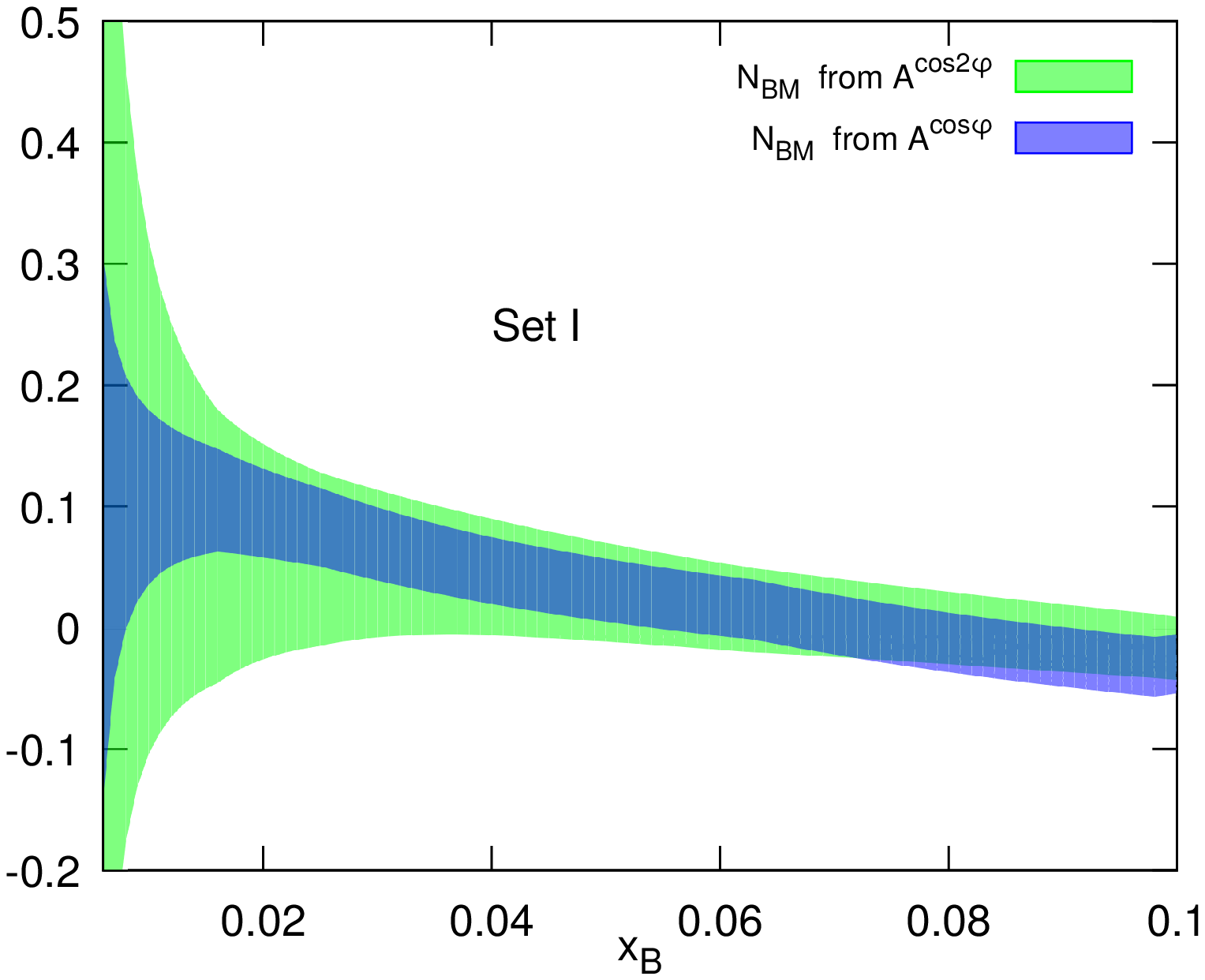}~~~\includegraphics[scale=0.3]{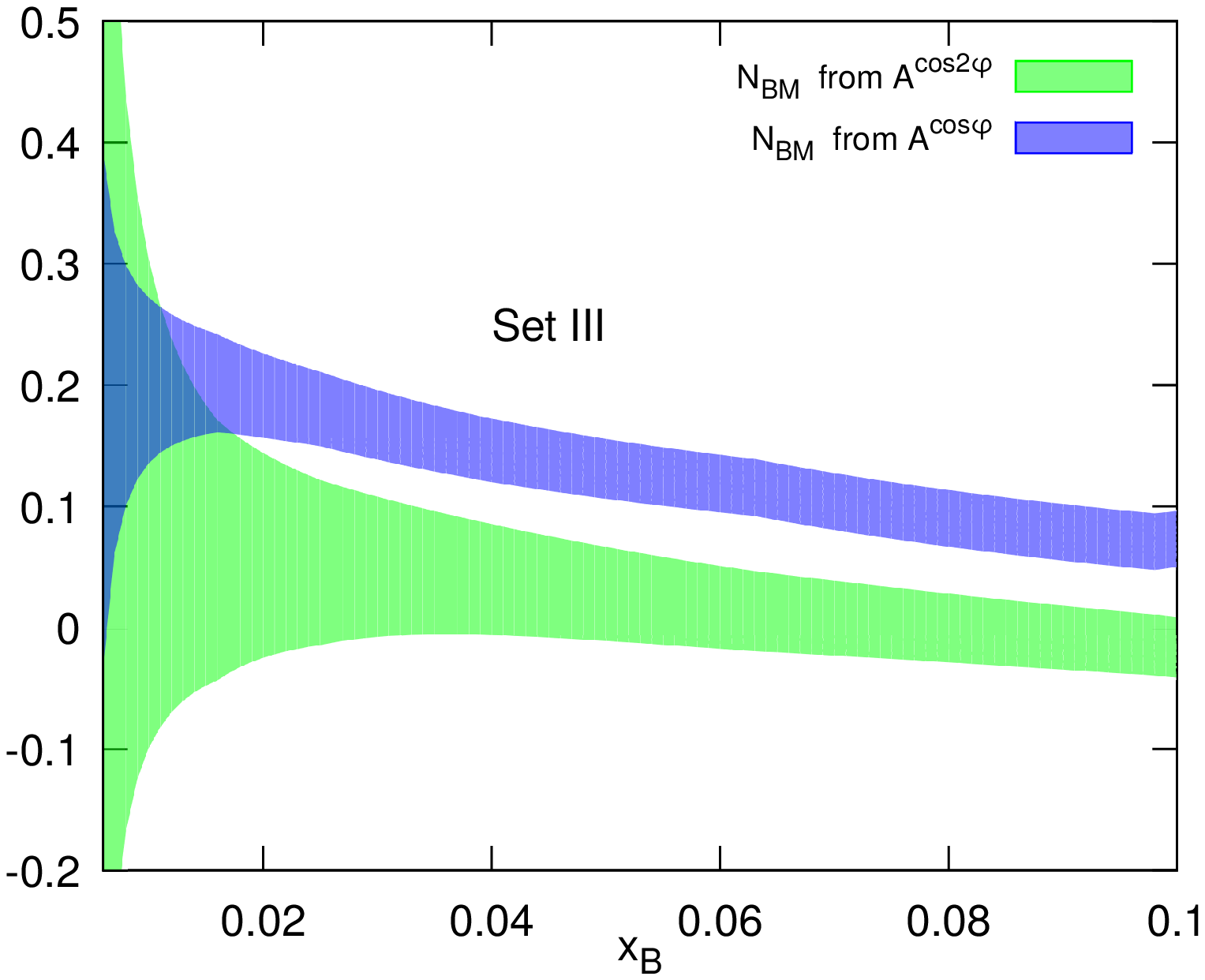}~~~\includegraphics[scale=0.3]{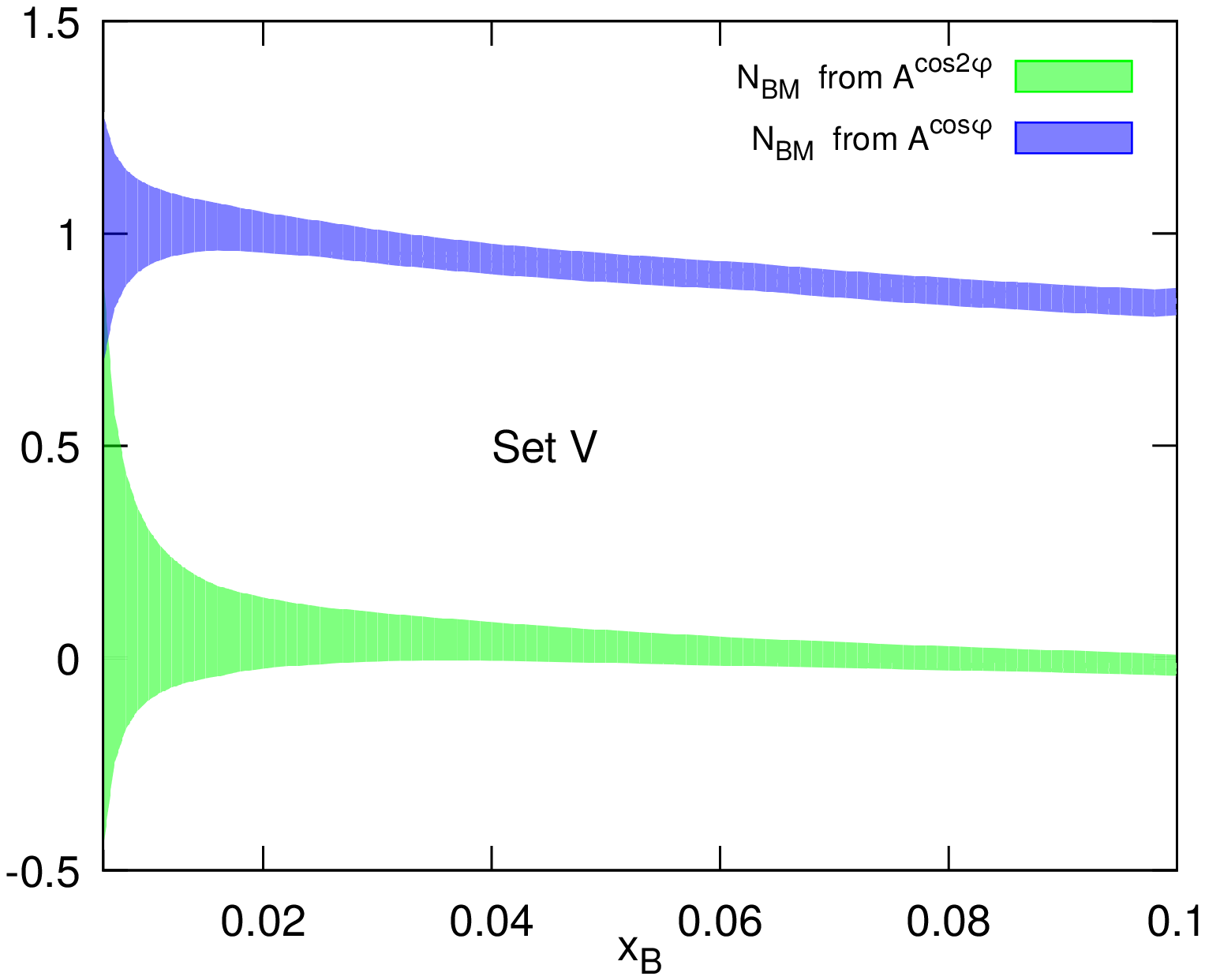}
\end{center}
\caption
{${\cal N}\BM^{Q_V}(\xb)$ extracted from the difference asymmetries, Eqs.~(\ref{A1}) and (\ref{A2}),
using different sets of parameters of Table~I and $\hat {\cal C}_{Cahn} + \hat {\cal C}_1$ instead
of $\hat {\cal C}_{Cahn}$. Again, plots for Sets. II and IV overlap with those for Sets. I and III, respectively.}
\label{fig:2}
\end{figure}

The relations  (\ref{A1}) and (\ref{A2}) provide 2 independent equations for the extraction of ${\cal N}\BM (\xb)$
for each set of the parameters in Table~I.
The results found in Fig.1 show that the 2 extractions are not completely compatible with each other for any choice
of the parameters given in Table~I. The source of the disagreement, we believe, lies in the value of the Cahn
contribution $\hat {\cal C}_{Cahn}$ in Eq.~(\ref{A2}). The point is that this Cahn term is a twist-4 contribution and there are
certainly other twist-4 contributions, from target mass corrections and other dynamic effects, which we  are unable to
calculate. One possibility would be to keep only twist-2 terms, but we think it interesting to obtain an estimate of
the missing twist-4 terms. We have therefore replaced $\hat {\cal C}_{Cahn}$ by
$\hat {\cal C}_{Cahn} + \hat {\cal C}_1$ , where $\hat {\cal C}_1$ is a free parameter adjusted to
improve the compatibility of the two extractions of ${\cal N}\BM^{Q_V}$  from Eqs.~(\ref{A1}) and (\ref{A2}).
We find perfect agreement for the parameter Set I  with  $\hat {\cal C}_{Cahn}$ replaced by
$\hat {\cal C}_{Cahn} + \hat {\cal C}_1$  [see Fig.~\ref{fig:2}] for the following parameter vaues:
 \be
 \avk =0.18,\quad\avp =0.20,\quad M\BM^2=0.34,\quad M\C^2=0.91,\quad\hat {\cal C}_1=-1.16. \label{best}
 \ee
Note that these  values for $\avk$ and $\avp$ agree with those obtained in \cite{Giordano:2008th}
and with the theoretical considerations \cite{Zavada:2009ska, Zavada:2011cv, DAlesio:2009cps}.
The  value obtained for $\hat {\cal C}_{Cahn} + \hat {\cal C}_1 = -0.85$
suggests that there are other twist-4 contributions, relatively large compared to the Cahn term,
in the $A^{\cos 2\phi ,h-\bar h}_{UU}$ asymmetry . \nl
An analytic expression for the extracted averaged ${\cal N}\BM^{Q_V}$ for the parameter Set given in
Eq.~(\ref{best}) is:
\be
{\cal N}\BM^{Q_V}(\xb ) &=& N\xb^\a (1-\xb)^\b (1+\g\xb), \nonumber \\
N = 0.475\pm 0.037, \quad \a &=& 0.242\pm 0.022, \quad \b = 13.3\pm 1.7, \quad \g = -13.7\pm 0.4\, .
\ee

\begin{figure}[H]
\begin{center}
\includegraphics[scale=0.4]{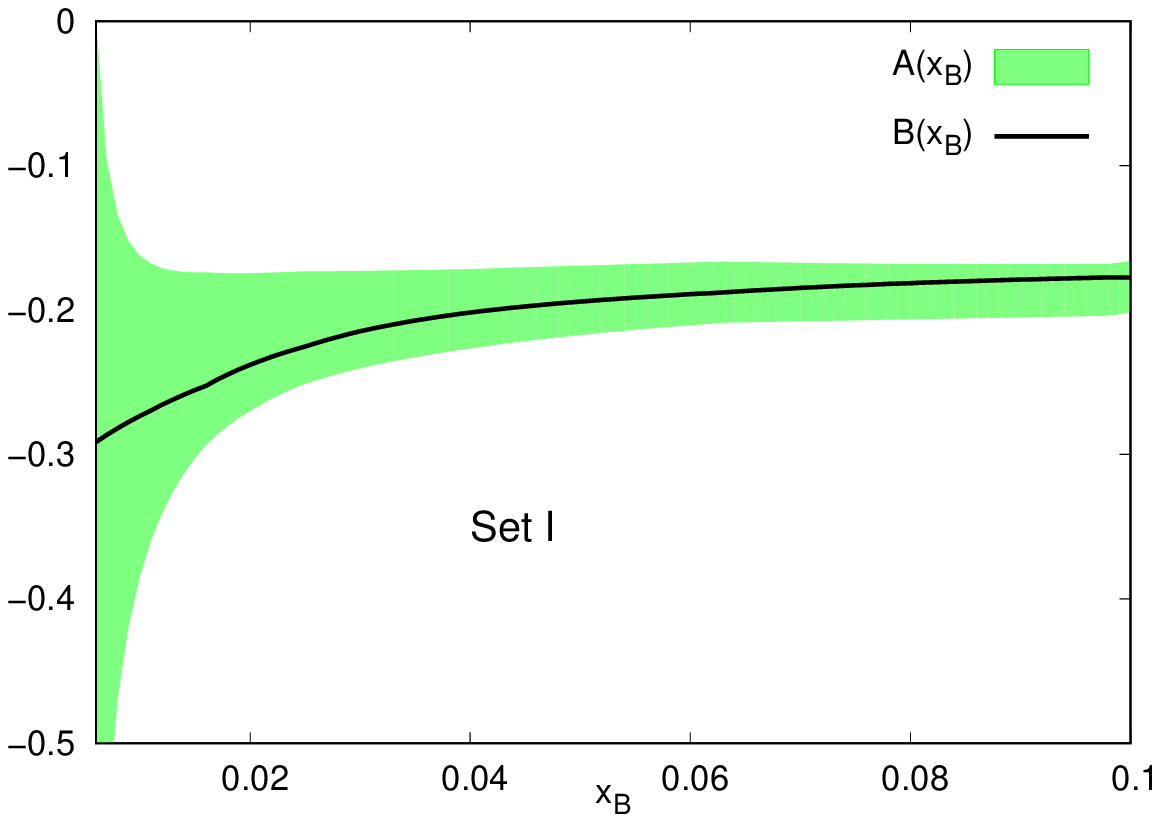}~~~\includegraphics[scale=0.4]{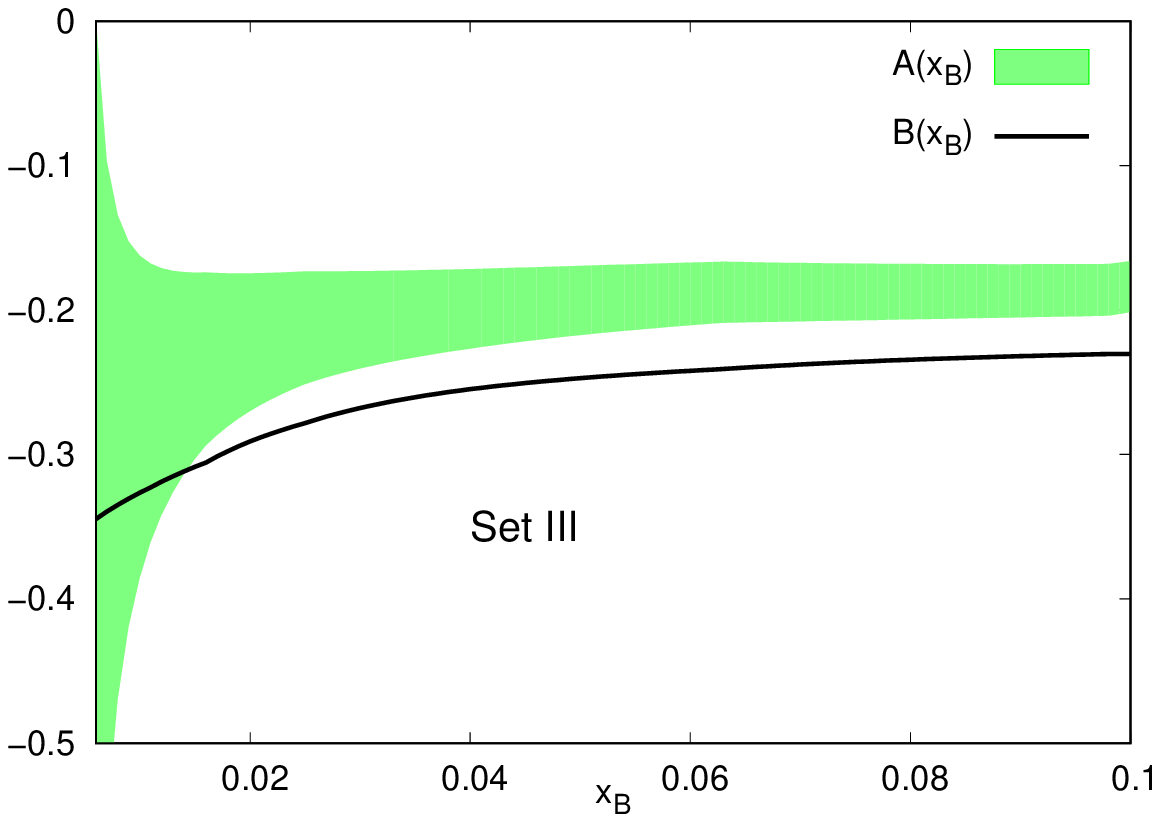}~~~\includegraphics[scale=0.4]{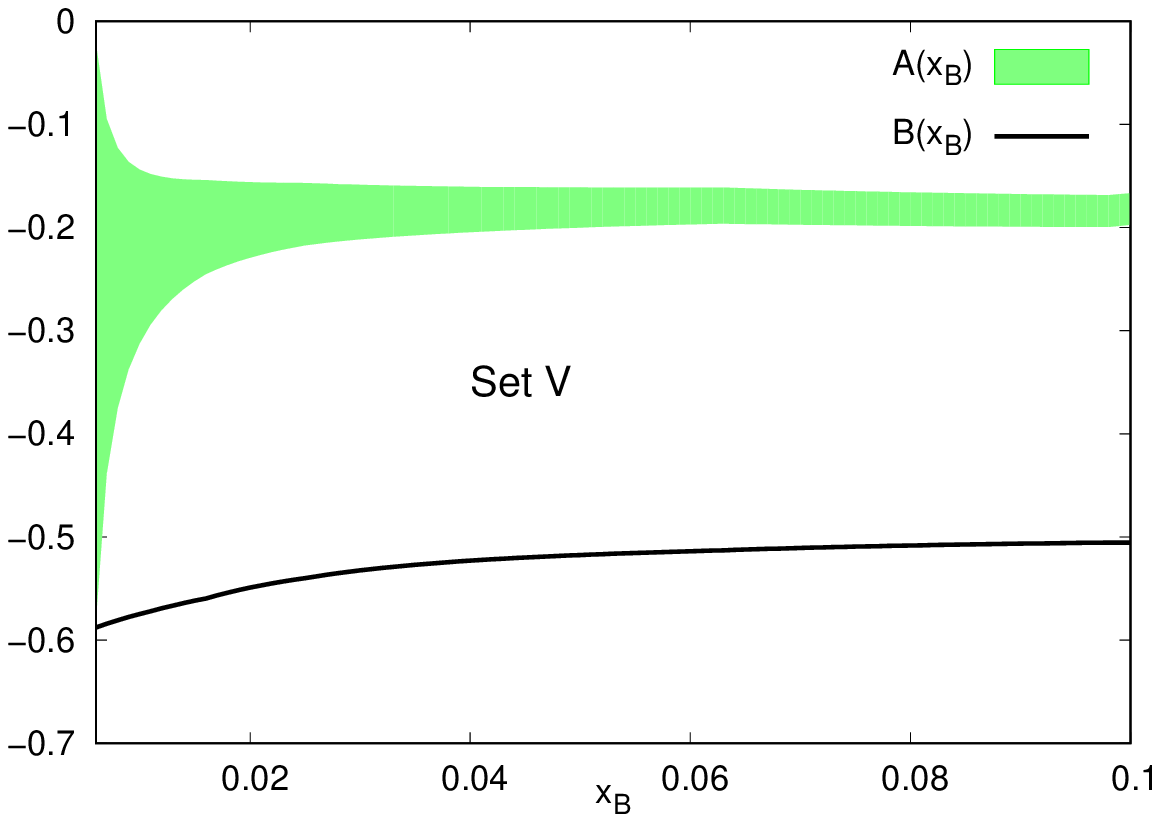}
\end{center}
\caption
{The test of Eq.~(\ref{rho}), with $\hat C_{Cahn}$  replaced by $\hat C_{Cahn} + \hat{C}_1$.
Again, plots for Sets. II and IV overlap with those for Sets I and III, respectively.}
\label{fig:3}
\end{figure}

Interestingly, there is a second way to utilize equations (\ref{A1}) and (\ref{A2}) which automatically imposes exact
consistency of the extraction of ${\cal N }\BM$, and which more directly fixes the values of the parameters
$\avk ,\, \avp\,, M\BM$, $M\C$  and  $\hat {\cal C}_1$. Eliminating ${\cal N}\BM^{Q_V}(\xb)$  from Eqs.~(\ref{A1})
and  (\ref{A2}) and using the variable $\rho$ we obtain:
\be
A(\xb )=B(\xb),\label{rho}
\ee
where
 \be
 A(\xb) &\equiv& \sqrt\frac{\langle Q^2\rangle(x_B)}{\avk}\,A_{UU,d}^{\cos\phi_h,h^+ -h^-}(\xb)\, +
\rho\,A_{UU,d}^{\cos2\phi_h,h^+ -h^-}(\xb),\\
 B(\xb) &\equiv& {\cal C}_{Cahn} + \rho\, \frac{\avk}{\langle Q^2\rangle(x_B)}\,\hat {\cal C}_{Cahn}.
 \ee
Fig.~\ref{fig:3} compares these two functions for various choices of the parameters
in Table~I. It is seen that there is  excellent agreement (with
$\hat {\cal C}_{Cahn}$  replaced by $\hat {\cal C}_{Cahn} + \hat {\cal C}_1$) for the values given in
Eq.~(\ref{best}).
We conclude therefore that the COMPASS data on $A_{UU}^{\cos  \phi_h}$ and $A_{UU}^{\cos  2\phi_h}$ strongly favour the
parameter values in Eq.~(\ref{best}).
This also confirms our suggestion that there are  significant twist-4 contributions other than the Cahn one.\\

\begin{figure}[H]
\begin{center}
\includegraphics[scale=0.5]{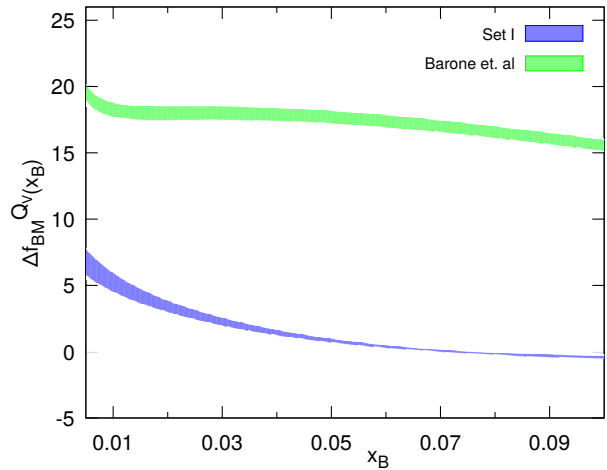}
\end{center}
\caption{Comparison of $\Delta f\BM^{Q_V}$ for Set I, Eq.~\ref{best}, with the result of Barone et al.\cite{Barone:2009hw}.
We use CTEQ6 parametrization for the collinear PDFs \cite{Pumplin:2002vw}.}
\label{fig:4}
\end{figure}

Our valence BM function $\Delta f\BM^{Q_V}(\xb)$ is shown in Fig.~\ref{fig:4},
where it is compared to $\Delta f\BM^{Q_V}(\xb)$ calculated from the
BM function published in \cite{Barone:2009hw}.
It is seen that there is a significant difference, suggesting that the BM
functions in \cite{Barone:2009hw} are incorrect. Note, as mentioned earlier, that the extraction in
\cite{Barone:2009hw} is, strictly speaking, theoretically inconsistent.
Finally we note that  future data on the $\langle\cos\phi_h\rangle$ and $\langle\cos 2\phi_h\rangle$ asymmetries on protons,  for charged pions or kaons, will allow access
to the BM function for the valence quarks  $u_V$ and $d_V$ separately, in the same essentially model independent manner
\cite{Christova:2015jsa}.

\section*{Acknowledgements}
E.C. and D.K. acknowledge the support of the Bulgarian-JINR collaborative Grant,
E.C. is grateful to Grant 08-17/2016 of the Bulgarian Science Foundation and
D.K. acknowledges the support of the Bogoliubov-Infeld Program.
D.K. thanks also A. Kotlorz for useful comments on numerical analysis.


\begin{thebibliography}{99}

\bibitem{Barone:2009hw} V. Barone, S. Melis and A. Prokudin,
Phys. Rev. D {\textbf 81}, 114026 (2010).

\bibitem{Barone:2008tn} V. Barone, A. Prokudin and Bo-Qiang Ma,
Phys. Rev. D {\textbf 78}, 045022 (2008).

\bibitem{Christova:2000nz} E. Christova and E. Leader, Nucl.Phys. B {\textbf 607}, 369 (2001).

\bibitem{Christova:2015jsa} E. Christova and E. Leader, Phys. Rev. D {\textbf 92}, 114004 (2015).

\bibitem{Adolph:2014pwc} C. Adolph \textit{et al.} (COMPASS Collaboration),
Nucl. Phys. B {\textbf 886}, 1046 (2014).

\bibitem{Anselmino:2011ch} M. Anselmino \textit{et al.}, Phys. Rev. D {\textbf 83}, 114019 (2011).

\bibitem{Christova:2014gva} E. Christova, Phys. Rev. D {\textbf 90} 054005 (2014).

\bibitem{Alekseev:2007vi} M. Alekseev \textit{et al.} (COMPASS Collaboration),
Phys. Lett B {\textbf 660}, 458 (2008).

\bibitem{Anselmino:2008jk} M. Anselmino \textit{et al.}, Nucl. Phys. B Proc. Suppl. \textbf{191}, 98 (2009).

\bibitem{Anselmino:2015sxa} M. Anselmino \textit{et al.}, Phys. Rev. D {\textbf 92}, 114023 (2015).

\bibitem{Anselmino:2015fty} M. Anselmino \textit{et al.}, Phys. Rev. D {\textbf 35}, 034025 (2016).

\bibitem{Christova:2017zxa} E. Christova, E. Leader and M.Stoilov, Phys. Rev. D {\textbf 97}, 056018 (2018).

\bibitem{DAlesio:2007bjf} U. D'Alesio and F. Murgia, Prog. Part. Nucl. Phys. {\textbf 61}, 394 (2008).

\bibitem{Bradamante:2007ex}
F. Bradamante, AIP Conf. Proc. 915, 513 (2007).

\bibitem{Albino:2008fy} S. Albino, B.A. Kniehl and G. Kramer, Nucl. Phys. B {\textbf 803}, 42 (2008).

\bibitem{Giordano:2008th} F. Giordano, report DESY-THESIS-2008-030

\bibitem{Zavada:2009ska} P. Zavada, Phys. Rev. D {\textbf 83}, 014022 (2011).

\bibitem{Zavada:2011cv} P. Zavada, Phys. Rev. D {\textbf 85}, 037501 (2012).

\bibitem{DAlesio:2009cps} U. D'Alesio, E. Leader and F. Murgia, Phys. Rev. D {\textbf 81}, 036010 (2010).

\bibitem{Pumplin:2002vw} J. Pumplin \textit{et al.}, J. High Energy Phys. 07 (2002) 012.

\end{thebibliography}
\end{document}